\documentclass[]{elsart}

\usepackage{graphicx}
\usepackage{amssymb}
\usepackage{epstopdf}
\usepackage{array}
\usepackage{subfigure}
\usepackage{url}
\usepackage{hyperref}
\usepackage{xspace}
\usepackage{lscape}
\usepackage{isorot}

\newtheorem{finding}{Finding}

\begin{document}

\begin{frontmatter}

\title{Towards slime mould colour sensor: Recognition of colours by \\ \emph{Physarum polycephalum}}
\author{Andrew Adamatzky}
\address{Unconventional Computing Centre, University of the West of England, Bristol, UK.  \url{andrew.adamatzky@uwe.ac.uk}}

\maketitle

\begin{abstract}
Acellular slime mould \emph{Physarum polycephalum} is a popular now user-friendly living substrate for designing
of future and emergent sensing and computing devices. \emph{P. polycephalum} exhibits regular patterns of 
oscillations of its surface electrical potential. The oscillation patterns are changed when the slime mould is subjected to 
mechanical, chemical, electrical or optical stimuli.  We evaluate feasibility of slime-mould based colour sensors by 
illuminating Physarum with red, green, blue and white colours and analysing patterns  of the  slime mould's 
electrical potential oscillations. We define that the slime mould recognises a colour if it reacts to illumination 
with the colour by a unique changes in amplitude and periods of oscillatory activity.  In laboratory experiments 
we found that the slime mould recognises red and blue colour. The slime mould does not differentiate between green 
and white colours. The slime mould also recognises when red colour is switched off.  We also map colours to diversity 
of the  oscillations: illumination with a white colour increases diversity of amplitudes and periods of oscillations, 
other colours studied increase diversity either of amplitude or period. 

\end{abstract}

\maketitle

\end{frontmatter}

\section{Introduction}

The plasmodium of \emph{Physarum polycephalum} (Order \emph{Physarales}, class \emph{Myxomecetes}, subclass \emph{Myxogastromycetidae}) is a single cell, visible with the naked eye, with many diploid nuclei. Plasmodium's foraging behaviour can be  interpreted as a computation~\cite{nakagaki_2000,nakagaki_iima_2007}:  data are represented by spatial of attractants and repellents, and  results are represented by structure of protoplasmic  network~\cite{adamatzky_physarummachines}.  Plasmodium can solve computational problems with natural parallelism, e.g. related to shortest path~\cite{nakagaki_2000} and hierarchies of planar proximity graphs~\cite{adamatzky_ppl_2008}, computation of plane tessellations~\cite{shirakawa}, execution of logical computing schemes~\cite{tsuda2004,adamatzky_gates}, and natural implementation of spatial logic and process algebra~\cite{schumann_adamatzky_2009}.  In the framework of our ``Physarum Chip'' EU project~\cite{adamatzky_phychip} we aim to experimentally implement a working prototype of a Physarum based general purpose computer. This computer will combine self-growing computing circuits made of a living slime mould with conventional electronic components. Data and control inputs to the Physarum Chip will be implemented via chemical, mechanical and optical means.

 Aiming to develop a component base of future Physarum computers we designed Physarum tactile sensor~\cite{adamatzky_2013_tactile} and
 undertook foundational studies towards fabrication of  slime mould chemical sensors (Physarum nose)~\cite{delacycostello_2013, whiting_2013}, and 
 uncovered memristive properties of the slime mould~\cite{gale_2013}.  Mechanical and, up to some degree, chemical stimulation could disturb
 structure of Physarum's protoplasmic networks and thus distort an architecture of a Physarum computer. Also massively parallel input of data into Physarum computer might be problematic. Implementation of inputs via optical means would be an ideal way of interaction with the slime mould computing devices. 
 
Plasmodium of \emph{P. polycephalum} shows a substantial degree of photo-sensitivity. A plasmodium moves away from 
light when it can or switches to another phase of its life cycle or undergoes fragmentation when it could not escape 
from light.  If a plasmodium, especially a starving one~\cite{guttes_1961}, is subjected to a high intensity of light 
the plasmodium turns into a sporulation phase~\cite{sauer_1969}. Phytochromes are involved in the 
light-induced sporulation~\cite{starostzik_1995}  and a sporulation morphogen is transferred by protoplasmic streams to 
all parts of the plasmodium~\cite{Hildebrandt_1986}. Photo-fragmentation is another physiological response to strong and unavoidable illumination. When a plasmodium is illuminated by ultraviolet or blue monochromatic light it breaks up into many equally sized fragments (each fragment contains around eight nuclei)~\cite{kakiuchi_2001}. The fragmentation is transient and after some time the fragments merge back into a fully functional plasmodium.

Photo-movement is a less (than sporulation or fragmentation) drastic response to illumination. Pioneer papers on photo-movement of Physarum reported that plasmodium exhibits the most pronounced negative photo-taxis to blue and white light~\cite{bailczyk_1979,schreckenbach_1980}. The illumination increase causes changes in the plasmodium's oscillatory activity; the degree of changes is proportional to the distance from the light source~\cite{Wohlfarth-Bottermann_1981,block_1981}. The exact mechanism of the response to light is as yet unknown. There are however a few phenomena uncovered in experiments. The first is the presence of phytochrome-like pigments~\cite{kakiuchi_2001}, which might be primary receptors of illumination. The light response of the pigments triggers a chain of biochemical processes~\cite{schreckenbach_1980}. These processes
include increase in activity of isomerase enzymes~\cite{starona_1992}, changes in mitochondrial respiration~\cite{korohoda_1983}
and spatially distributed oscillations in ATP concentrations~\cite{ueda_1986}.

Nakagaki et al.~\cite{nakagaki_1999,nakagaki_iima_2007} undertook the first ever experiments on shaping plasmodium
behaviour with illumination.  They discovered that protoplasm streaming oscillations of  plasmodium can be tuned by,
or relatively synchronised with, periodic illumination~\cite{nakagaki_1999}. They also demonstrated that plasmodium optimises its protoplasmic network structure in a field with heterogeneous illumination~\cite{nakagaki_iima_2007}: the thicknesses of protoplasmic tubes
in illuminated areas  are less than the thicknesses of tubes in shaded areas~\cite{nakagaki_iima_2007}. Inspired by our previous experiments on 
routing active growing zones of Physarum with localised domains of illumination~\cite{adamatzky_2009} we decided to investigate what would be a
fine reaction of Physarum to illumination with different colours. 

The reaction of Physarum to colour of illumination was measured via recording of 
patterns of Physarum's electrical activity. An undisturbed Physarum exhibits more or less regular patterns of oscillations of its surface electrical potential. The electrical potential oscillations are more likely controlling a peristaltic activity of protoplasmic tubes, necessary for distribution of nutrients in the spatially extended body of Physarum~\cite{seifriz_1937,heilbrunn_1939}. A calcium ion flux through membrane triggers oscillators responsible for dynamic of contractile activity~\cite{meyer_1979,fingerle_1982}.  The potential oscillates with amplitude of 1 to 10~mV and period 30-200~sec, associated with shuttle streaming of cytoplasm~\cite{iwamura_1949, kamiya_1950, kashimoto_1958, meyer_1979}.  In our experiments we observed sometimes lower amplitudes because there are agar blobs between Physarum and electrodes and, also, recording and references electrodes were connected with each other via a protoplasmic tube only. Exact characteristics of electric potential oscillations vary depending on state of Physarum culture and experimental setups~\cite{achenbach_1980,achenbach_1981}.

 We define that the slime mould recognises a colour if it reacts to illumination with the colour by a unique changes in amplitude and periods of oscillatory activity. We aim to answer the question: Does plasmodium of \emph{P. polycephalum} recognise red, green, blue and white colours? 

\section{Methods}
\label{methods}

\begin{figure}[!tbp] 
\centering
\includegraphics[width=0.6\textwidth]{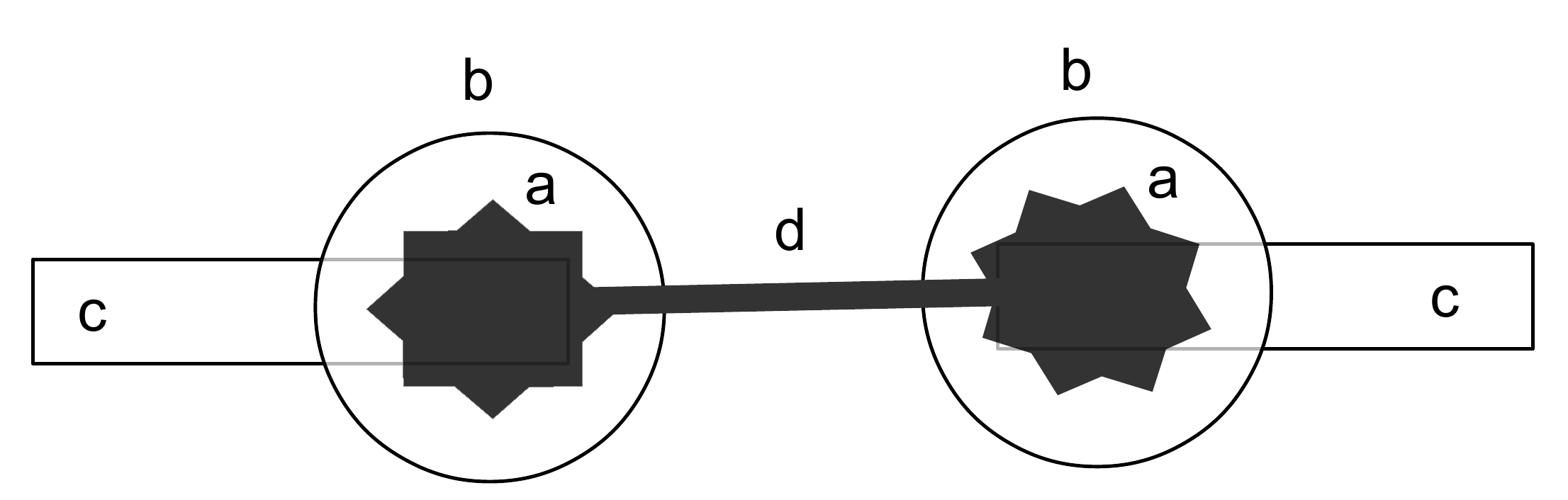}
\caption{A scheme of experimental setup: (a)~Physarum, (b)~agar blobs, (c)~electrodes, (d)~protoplasmic tube. All parts of Physarum 
shown in dark grey form a single cell.}
\label{scheme}
\end{figure}

\begin{figure}[!tbp] 
\centering
\subfigure[]{\includegraphics[width=0.6\textwidth]{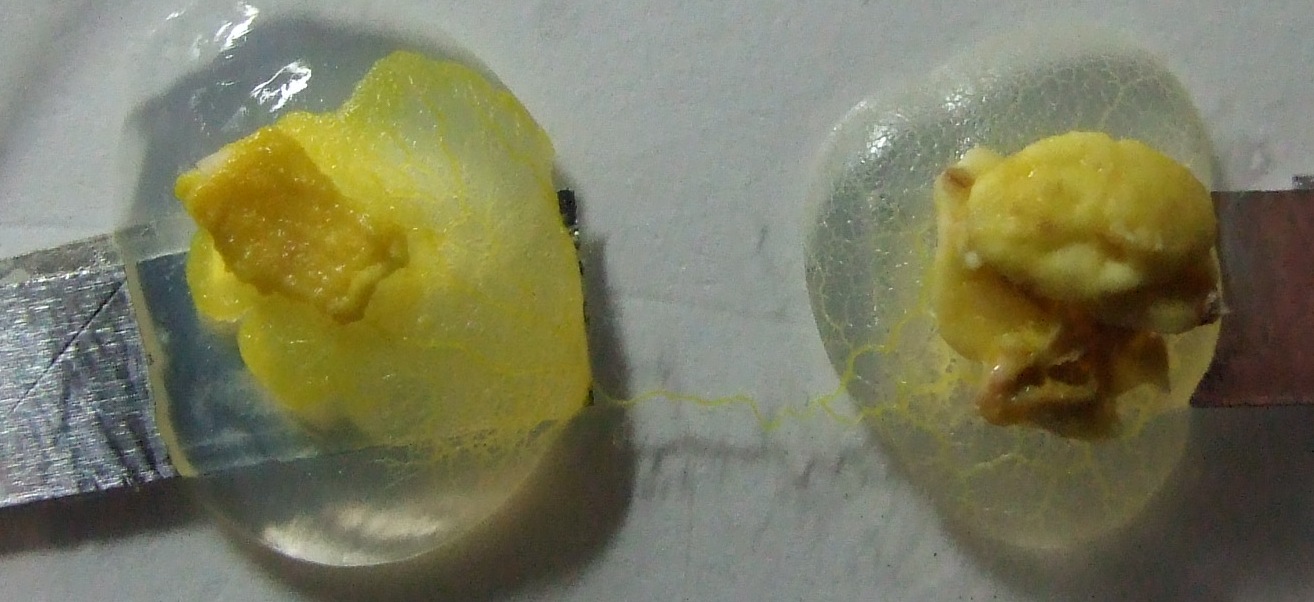}}
\subfigure[]{\includegraphics[width=0.6\textwidth]{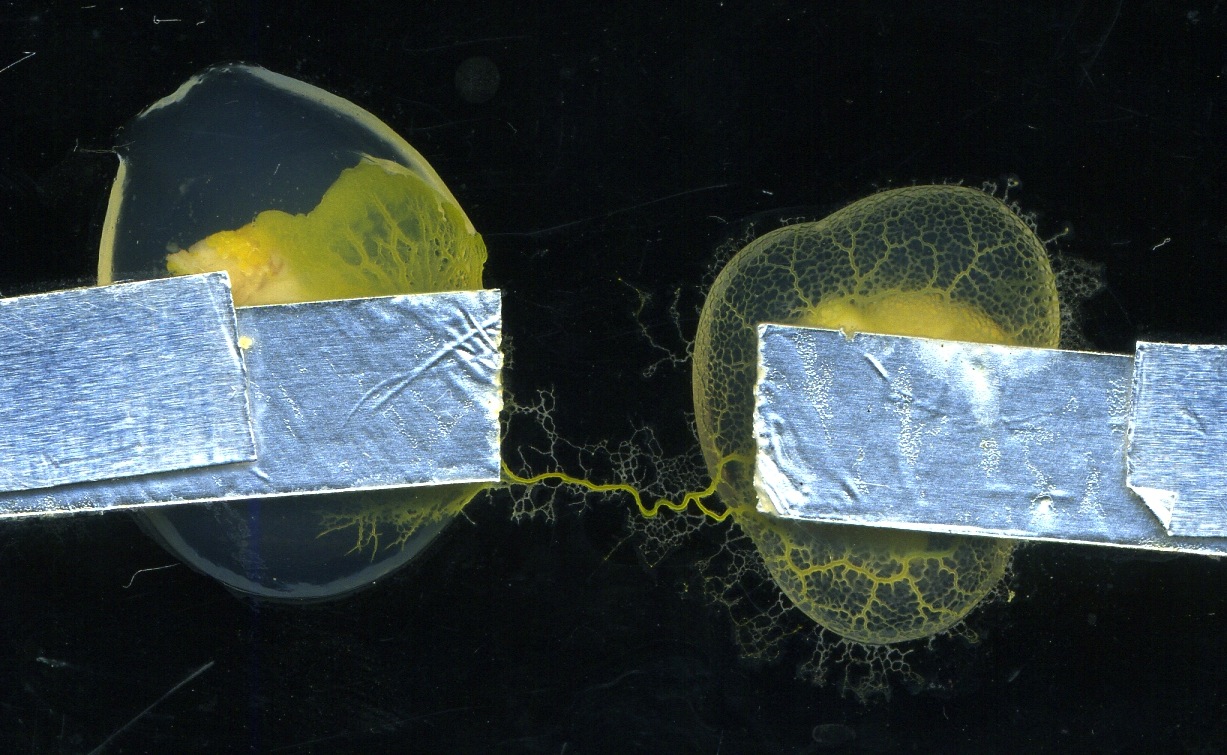}}
\caption{Protoplasmic tube connects two blobs of agar lying on electrodes. (a)~Photo from above. (b)~Scan of Petri dish from below.}
\label{photos}
\end{figure}

\begin{figure}[!tbp] 
\centering
\subfigure[]{\includegraphics[width=0.8\textwidth]{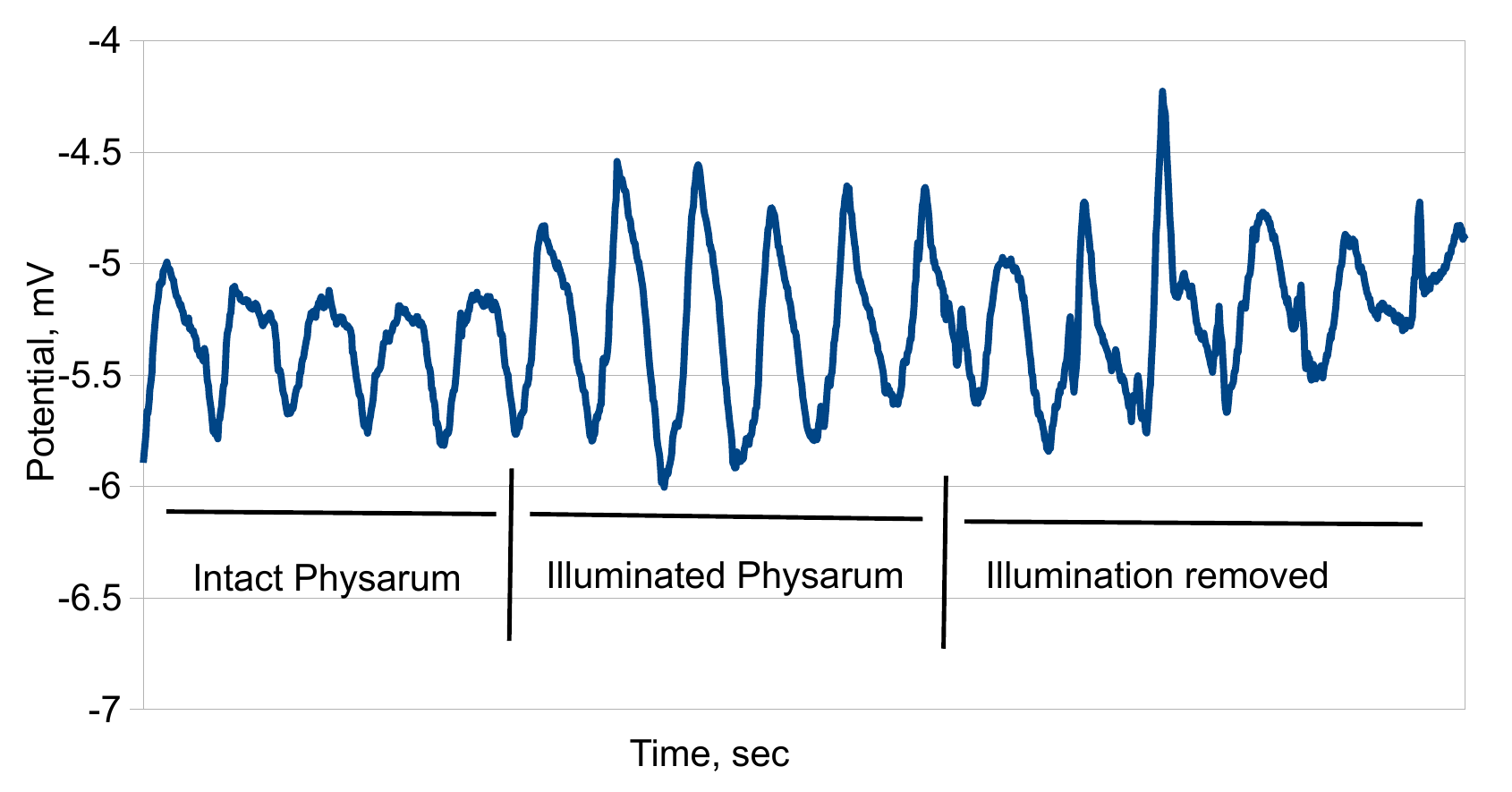}}
\subfigure[]{\includegraphics[width=0.8\textwidth]{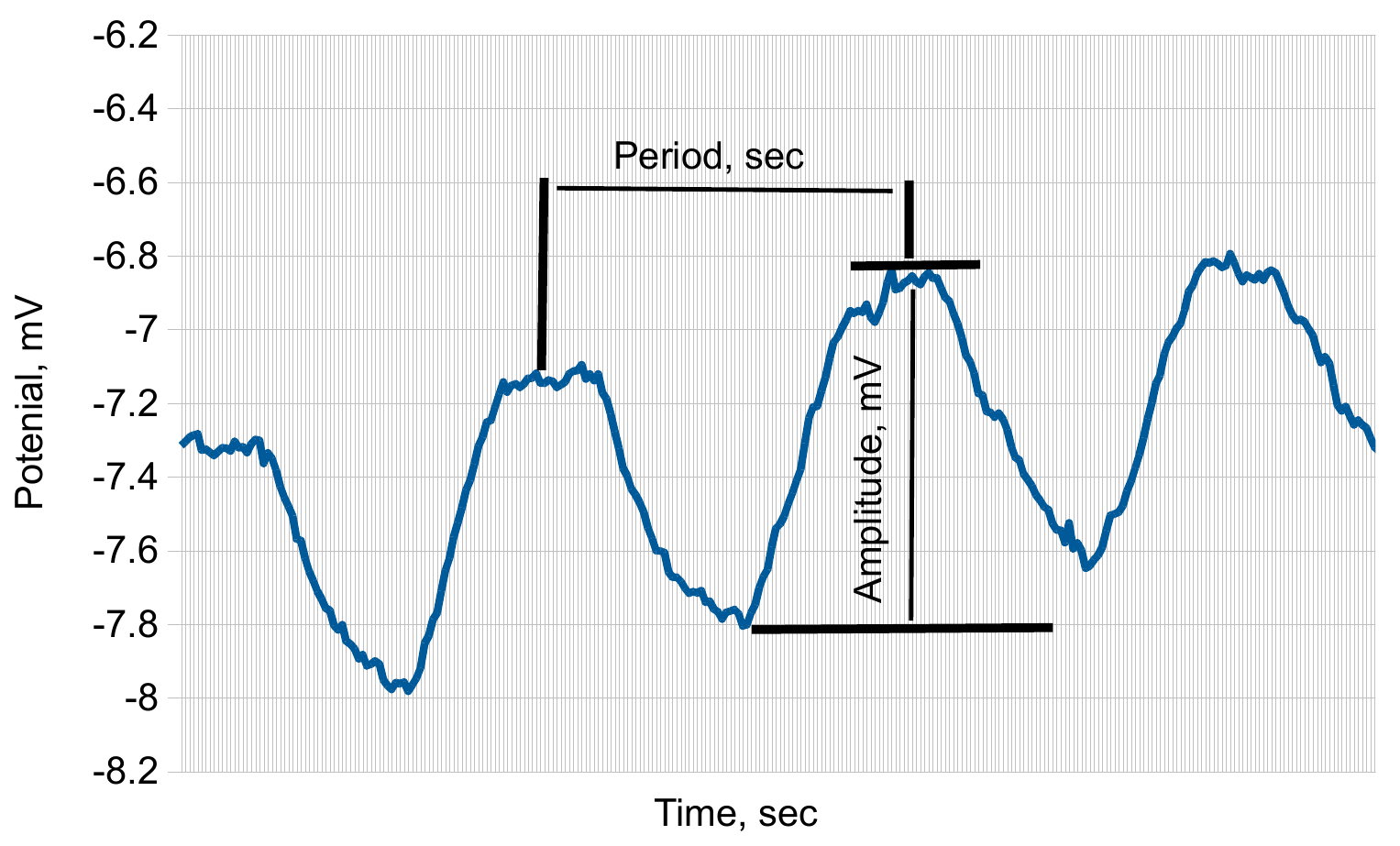}}
\caption{Patterns of oscillations of Physarum surface electrical potential. 
(a)~Example of oscillations before, during and after illumination. 
(b)~Scheme of measurements taken.}
\label{measurements}
\end{figure}

Plasmodium of \emph{Physarum polycephalum} was cultivated in plastic boxes on wet kitchen towels and fed with oat flakes. Culture was periodically replanted to a fresh substrate. Electrical activity of plasmodium was recorded with  
ADC-20 High Resolution Data Logger  (Pico Technology, UK).  The data logger ADC-24 employs differential inputs, galvanic isolation and software-selectable sample rates all contribute to a superior noise-free resolution; its 24-bit A/D converted maintains a gain error of 0.1\%. Its input impedance is 2~M$\Omega$ for differential inputs, and offset error is 36~$\mu$V in $\pm$ 1250~mV range use

A scheme of experimental setup is shown in Fig.~\ref{scheme} and photographs are
shown in Fig.~\ref{photos}. 
Each electrode is made of a conductive aluminium foil, 0.07~mm thick,  8~mm wide, 50~mm (inclusive part protruding outside Petri dish) long.
Two blobs of agar 2 ml each (Fig.~\ref{scheme}b) were placed on electrodes (Fig.~\ref{scheme}c) stuck  to a bottom of a plastic Petri dish (9~cm). 
Distance between proximal sites of electrodes is always 10~mm. Physarum was inoculated on one agar blob. We waited till Physarum colonised the first blob, where it was inoculated, and propagated towards and colonised the second blob. When second blob is colonised, two blobs of agar, both colonised by Physarum (Fig.~\ref{scheme}a), became connected by a single protoplasmic tube  (Fig.~\ref{scheme}d). We discounted experiments more than one tube was formed between the blobs because because patterns of oscillation were affected by interactions between potential waves travelling along interlinked protoplasmic tubes. Petri dished were kept in darkness before and during recordings. This experimental setup is proved to be reliable in studies of electrical activity of Physarum~\cite{adamatzky_jones_2011,adamatzky_2013_tactile}.

A blob based on recording electrode was illuminated from above using using white LED 1400 LUX with a set of colour lenses: red (635~nm), green (560~nm) and blue (450~nm).  We also illuminated Physarum with white light via transparent lens. We adjusted heigh of source of light (typically circa
15-17~mm above agar blob) such that the whole agar blob, c. 10~mm diameter, was illuminated.   Amount of light on the blob was 80-120 LUX for each colour.  In each experiment we recorded electrical activity of Physarum in darkness (10~min), under illumination (10~min) and after illumination was removed (10~min), see Fig.~\ref{measurements}a. For each colour we conducted 30 experiments. 

 For each recording and type of illumination we calculated average period, average amplitude and standard deviations of these two 
 (Fig.~\ref{measurements}b). Average amplitudes and periods of oscillation of non-illuminated Physarum ($A$ and $P$), Physarum under illumination  ($A'$ and $P'$) and Physarum after illumination was switched off ($A''$ and $P''$).
Diversity of electrical activity was estimated via standard deviations of amplitudes and periods $\sigma A$, $\sigma P$,
$\sigma A'$, $\sigma P'$, $\sigma A''$, $\sigma P''$. The standard deviations were calculated during each experiment.

\section{Results}

\begin{table}[!tbp]
\caption{Average amplitudes $A$, $A'$, $A''$ and average periods of oscillations of Physarum surface potential measured in 30 experiments for each type of illumination: $A$ and $P$ are average amplitude and period before illumination, $A'$ and $P'$ during illumination, $A''$ and $P''$ after 
illumination was switched off. Amplitudes $A$, $A'$ and $A''$ are measured in mV, and periods $P$, $P'$ and $P''$ in seconds.}
\begin{scriptsize}
\begin{tabular}{l|l|llllllllllll}
Colour	& Value		&	$A$	&	$\sigma A$	&	$A'$	&	$\sigma A'$	&	$A''$	&	$\sigma A''$	&	$P$	&	$\sigma P$	&	$P'$	&	$\sigma P'$	&	$P''$	&	$\sigma P''$	\\ \hline
Red	&	Average	&	0.7	&	0.26	&	0.8	&	0.31	&	0.72	&	0.23	&	104.77	&	19.88	&	105.62	&	27.74	&	151.76	&	33.19	\\
	&	Stn. deviation	&	0.27	&	0.21	&	0.25	&	0.18	&	0.33	&	0.24	&	33.95	&	15.02	&	23.07	&	9.57	&	180.31	&	17.99	\\
Green	&	Average	&	1.55	&	0.34	&	1.23	&	0.64	&	1.12	&	0.32	&	117.18	&	16.71	&	110.22	&	24.37	&	100.49	&	25.21	\\
	&	Stn. deviation	&	0.8	&	0.23	&	0.59	&	0.38	&	0.62	&	0.33	&	18.84	&	12.54	&	23.31	&	10.92	&	21.44	&	14.74	\\
Blue	&	Average	&	0.78	&	0.25	&	0.76	&	0.31	&	0.68	&	0.24	&	94.33	&	17.46	&	110.62	&	29.66	&	104.2	&	30.07	\\
	&	Stn. deviation	&	0.4	&	0.17	&	0.29	&	0.25	&	0.3	&	0.18	&	23.91	&	11.29	&	33.56	&	9.87	&	34.68	&	12.05	\\
White	&	Average	&	0.69	&	0.28	&	0.75	&	0.37	&	0.65	&	0.23	&	86.44	&	23.12	&	79.54	&	27.75	&	78.76	&	26.3	\\
	&	Stn. deviation	&	0.35	&	0.19	&	0.37	&	0.19	&	0.38	&	0.17	&	18.56	&	7.19	&	15.28	&	10.29	&	19.03	&	8.33	\\
\end{tabular}
\end{scriptsize}
\label{experimentalvalues}
\end{table}

\begin{table}[!tbp]
\caption{Relative changes in amplitudes, periods and diversities of amplitudes and periods: 
$A'/A$, $P'/P$ and $\sigma A'/\sigma A$, $\sigma P'/\sigma P$ (after illumination is switched on), 
$A''/A'$, $P''/P'$ and $\sigma A''/\sigma A'$, $\sigma P''/\sigma P'$ (after illumination is switched off). }
\begin{scriptsize}
\begin{tabular}{l|l|llllllllllll}
Colour	& Value		&	$A'/A$	&	$A''/A'$	&	$P'/P$	&	$P''/P'$	&	$\sigma A'/\sigma A$	&	$\sigma A''/\sigma A'$	&	$\sigma P'/\sigma P$	&	$\sigma P''/\sigma P'$	\\ \hline
RED	&	Average	&	1.08	&	0.74	&	1.01	&	1.04	&	0.64	&	0.42	&	1.35	&	0.86	\\
	&	Stn. Deviation	&	2.38	&	1.23	&	4.17	&	2.78	&	0.28	&	0.48	&	1.72	&	1.08	\\
GREEN	&	Average	&	0.68	&	0.77	&	0.94	&	0.88	&	1.49	&	0.22	&	0.89	&	0.68	\\
	&	Stn. Deviation	&	1.27	&	1.32	&	4.76	&	3.03	&	2.08	&	0.21	&	1.02	&	0.7	\\
BLUE	&	Average	&	0.85	&	0.79	&	1.12	&	0.91	&	0.92	&	0.39	&	1.56	&	0.93	\\
	&	Stn. Deviation	&	1.1	&	1.85	&	4.76	&	3.85	&	0.98	&	0.2	&	2.38	&	2.44	\\
WHITE	&	Average	&	0.98	&	0.73	&	0.89	&	0.95	&	1.14	&	0.43	&	1.04	&	0.87	\\
	&	Stn. Deviation	&	2.17	&	1.3	&	3.45	&	3.57	&	1.49	&	0.51	&	1.72	&	1.92	\\
\end{tabular}
\end{scriptsize}
\label{relativevalues}
\end{table}

Absolute parameters of potential oscillations are shown in Tab.~\ref{experimentalvalues}. Electrical behaviour of Physarum shows high degree of variability. Oscillations of intact Physarum vary from 0.69~mV to 1..55~mV, averaged 0.93~mV over all 120 trials. Fastest oscillations of non-illuminated Physarum occur with period c. 86~sec and slowest oscillations with intervals of almost two minutes between waveforms.  Oscillation amplitudes of illuminated Physarum range from 0.75~mV to  1.23~mV, averaged 0.88~mV, and periods from c.79~sec to 110~sec. When illumination is switched off
surface electrical potential of Physarum oscillates with amplitudes from 0.65~mV to 1.12~mV, averaged 0.79~mV.  Intervals between waveforms of 
post-illumination Physarum range from 79~sec to 151~sec (Tab.~\ref{experimentalvalues}).

Due to high variability of oscillation patterns it would be unwise to compare absolute values of oscillation amplitudes and periods. Oscillation patterns of 
intact and undisturbed Physarum can differ substantially between trials. It is rather reasonable to consider relative values of changes in oscillation patterns in each experimental trial: after illumination is applied ($A'/A$, $P'/P$, $\sigma A' /\sigma A$, $\sigma P' /\sigma P$) and
after illumination is switched off  ($A''/A'$, $P''/P'$, $\sigma A'' /\sigma A'$, $\sigma P'' /\sigma P'$). 
These relative values are shown in Tab.~\ref{relativevalues}.  Illumination does not modify patterns of oscillation substantially. Largest decrease in 
amplitude is 32\% of intact Physarum's amplitude. Such decrease happens when Physarum is illuminated by  green colour.  Largest increase in amplitude, 108\%, is caused by red illumination. Most substantial decreases in periods of oscillation, 11-12\%, are caused by illumination of Physarum with white light or switching off green illumination. 

Disregarding a colour of illumination we observe that amplitude of electrical potential oscillations --- averaged over all trials --- is 0.9 of an amplitude of a non-illuminated Physarum. When illumination is switched of an average becomes even smaller, just 0.75 of an amplitude of illuminated Physarum.  
Average periods of oscillations after switching illumination on or off are decrease just by 1\% or 6\%. Thus, averaging over all colours of illumination, amplitude of oscillations usually decreases and frequency slightly increases when illumination is switched on or off.     
Indeed, as we discuss below, effects of different colours may be distinctive.

\begin{figure}[!tbp] 
\centering
\subfigure[Illumination is ON]{\includegraphics[width=0.45\textwidth]{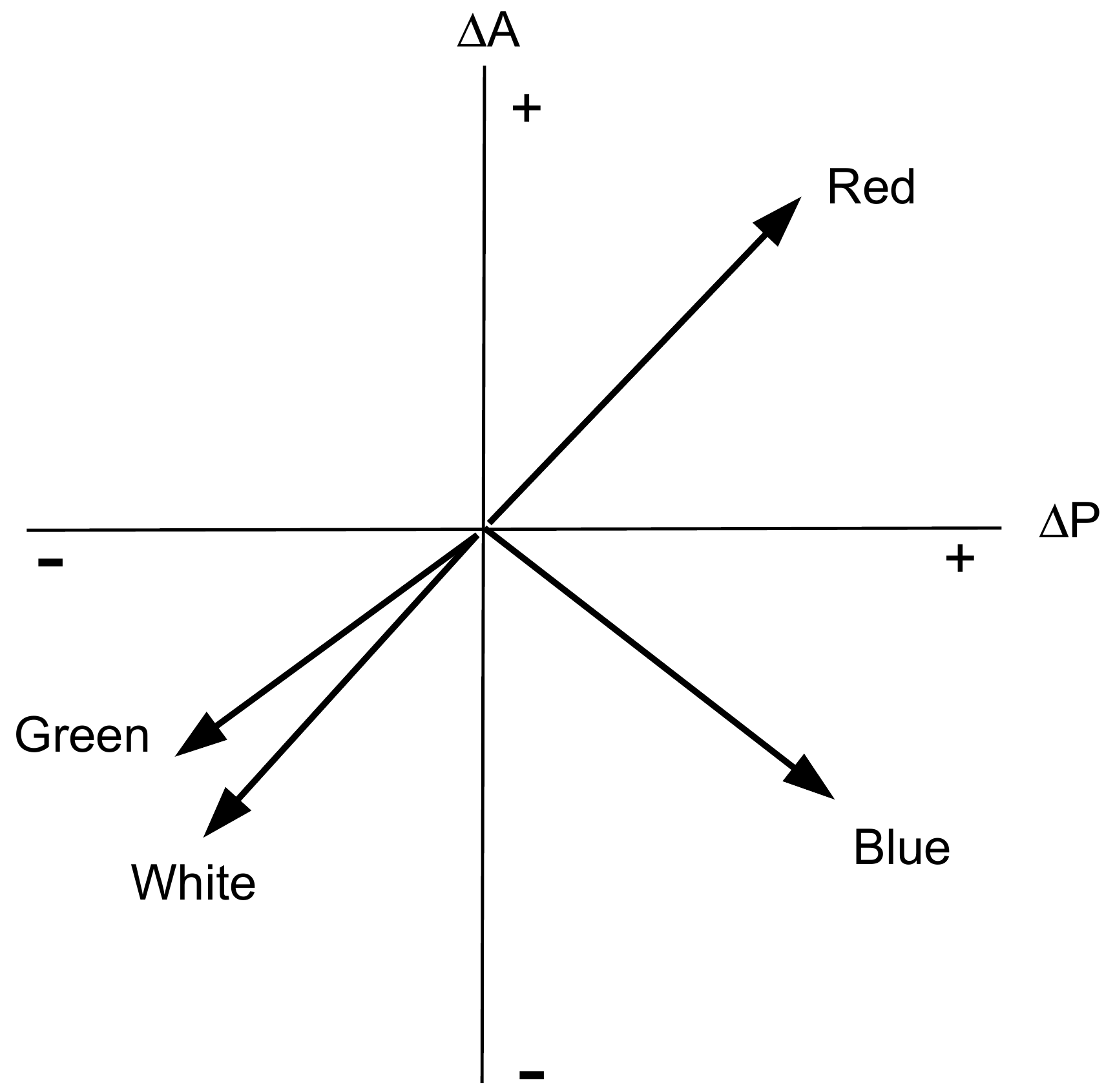}}
\subfigure[Illumination is OFF]{\includegraphics[width=0.45\textwidth]{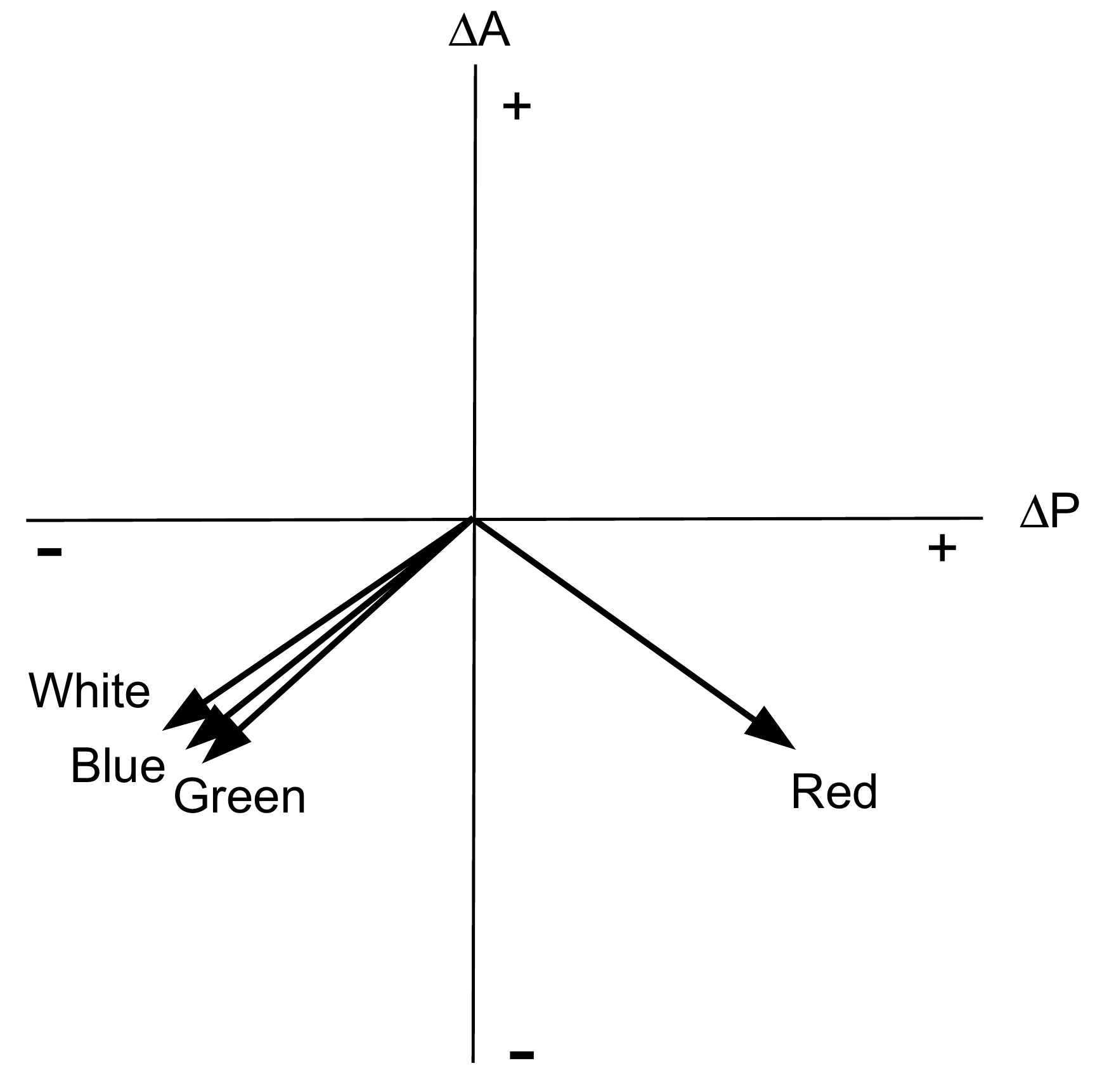}}
\subfigure[Illumination is ON]{\includegraphics[width=0.45\textwidth]{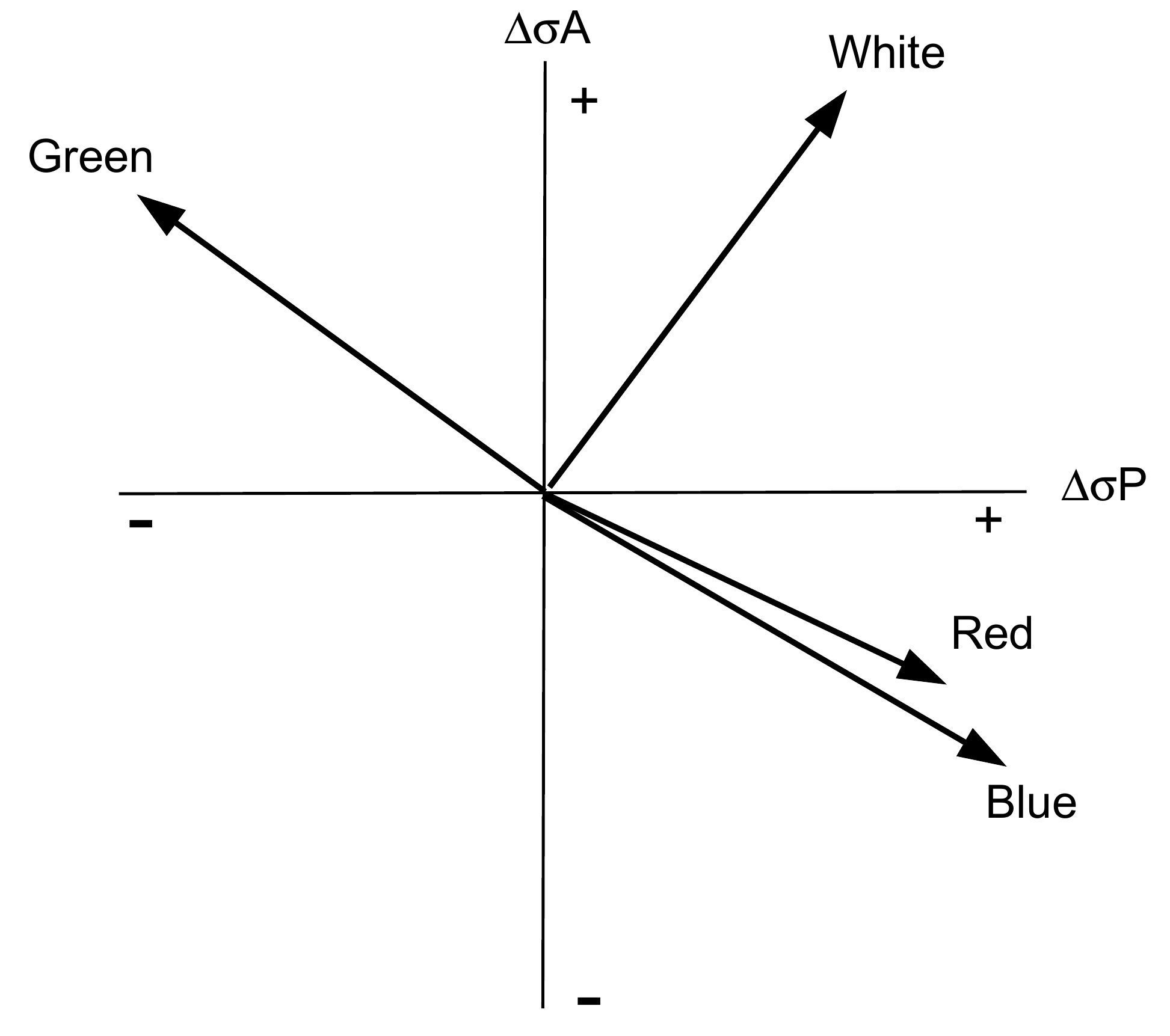}}
\subfigure[Illumination is OFF]{\includegraphics[width=0.45\textwidth]{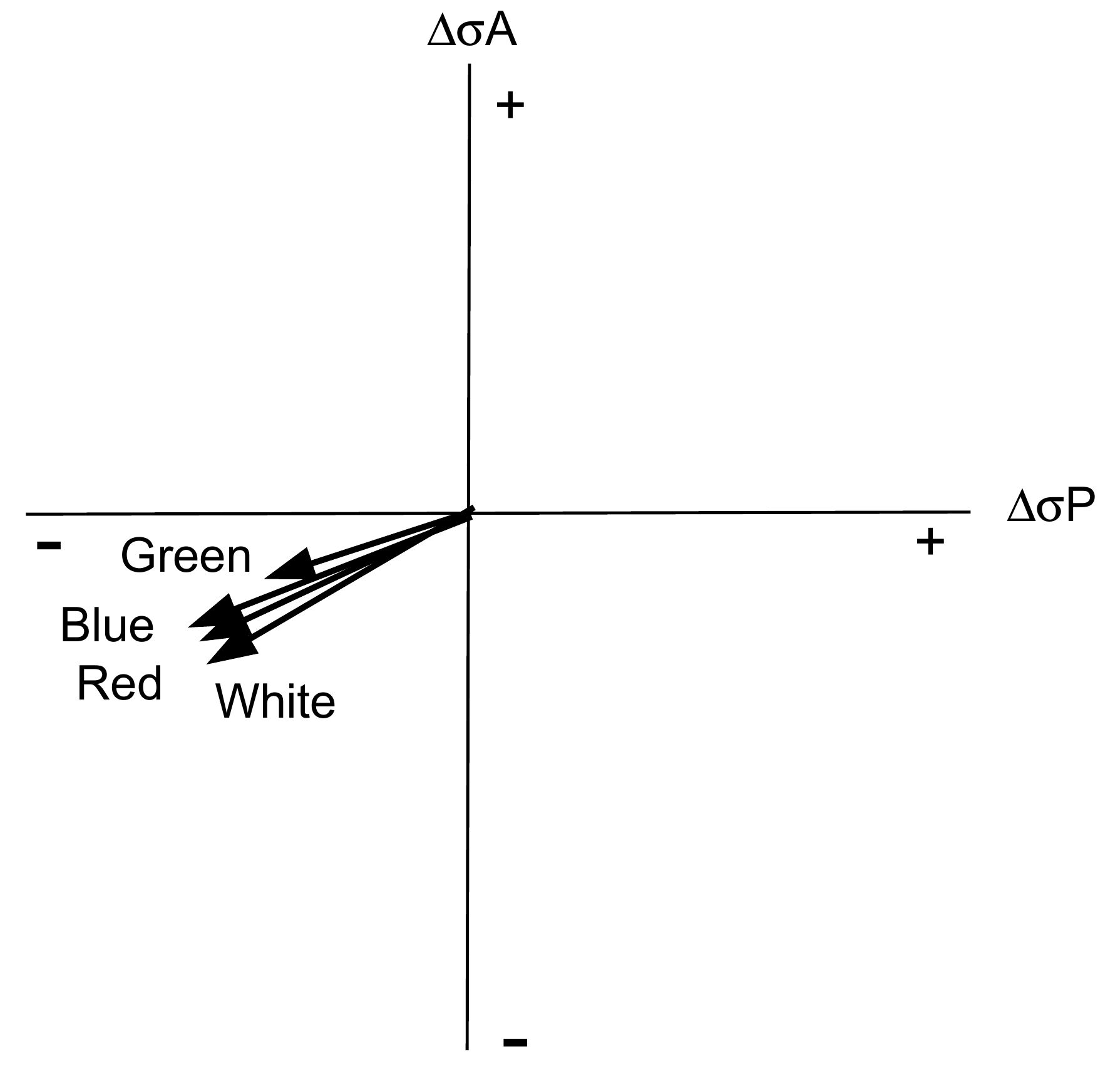}}
\caption{Scheme of changes $\Delta w$, where $w$ is $A$, $A'$, $P$, $P'$, $\sigma A$, $\sigma A'$, $\sigma P$ and $\sigma P'$ 
induced by colours of illumination.  Lengths of vectors are calculated from Tab.~\ref{relativevalues} as $\Delta w = \frac{w^*}{w}-1$, where 
$w$ is a parameter before stimulation and $w^*$ after stimulation.}
\label{schemes}
\end{figure}

 We define that the slime mould recognises a colour $c$ if it reacts to illumination with the colour $c$ by a unique changes in amplitude and periods of oscillatory activity. Let $w$ be a parameter (average amplitude, average period, standard deviations of amplitudes or periods) 
of oscillations before stimulus applied or removed, and $w^{*}$  after stimulus applied or removed. Then we 
assume $\Delta w = \frac{w^*}{w}-1$.  Graphical representation  of $\Delta w$ is given in Fig.~\ref{schemes}.

\begin{finding}
Physarum recognises when red and blue light are switched on and when red light is 
switched off.
\end{finding}

Red and blue occupy their own quadrants in $\Delta A$--$\Delta P$ while green and white lights share the same quadrant (Fig.~\ref{schemes}a).
Red and blue illuminations decrease frequency of oscillations, i.e. increase period. 
Red light increases amplitude of oscillations but blue light decreases the amplitude. Physarum does not differentiate between
green and white lights.  Switching off all lights but red both amplitude and frequency of oscillations  (Fig.~\ref{schemes}b).
Switching off red light leads to increase of periods and decrease of amplitudes of oscillations  (Fig.~\ref{schemes}b).

Diversity of oscillations, calculated as a standard deviation of amplitudes or periods, is another useful characteristics of Physarum response to 
illumination. 

\begin{finding}
In terms of diversity of oscillations, Physarum recognises when white and green colours are switched on. 
\end{finding}

Physarum's responses, in terms of diversity of oscillations, to red an green light are in  their unique quadrants on 
$\Delta \sigma A$--$\Delta \sigma P$  (Fig.~\ref{schemes}c). Red and blue colours share the same quadrant with 
each other. Red and blue lights increase diversity of periods and decreases diversity of amplitudes. Green light  increases 
diversity of amplitudes and decreases diversity of periods. White light increases diversity of amplitudes and periods.

\begin{finding}
In terms of diversity of oscillations,  Physarum recognises that illumination is switched off (its oscillating behaviour becomes uniform) 
but does not recognise what exact colour is switched off. 
\end{finding}

Switching off all types of illumination decreases diversity of amplitudes and periods (Fig.~\ref{schemes}d).

\section{Discussion}

In laboratory experiments with slime mould \emph{P. polycephalum} we demonstrated that the slime mould responds to illumination with red and blue lights with unique patterns of oscillation of its surface electrical potential. Physarum does not differentiate between green and white lights. Switching illumination off also modifies characteristics of Physarum potential oscillations: the slime mould differentiate between events when red light was switched off and when all other lights (green, blue, white) were off. In experiments we also considered a diversity of oscillations. We found that, in terms of diversity of oscillations,  Physarum reacts to illumination by increasing diversity of amplitudes (white and green lights) or periods (red, blue and white lights) of oscillations. Increase in diversity of oscillation might be explained by formation of additional micro-oscillator in Physarum protoplasmic networks. Phases and frequencies of oscillations and positions of micro-oscillators relative  to each other lead to emergence of waveforms with different amplitudes and periods, as recorded in experiments. Switching off illumination may extinguish some of  the micro-oscillators, therefore Physarum reacts to switching off the illumination by producing rather uniform patterns of oscillation. Our results well advance previous 
studies on photo-sensitivity of the slime mould~\cite{bailczyk_1979, schreckenbach_1980, block_1981, Wohlfarth-Bottermann_1981, korohoda_1983, starona_1992, starostzik_1995, nakagaki_1999, kakiuchi_2001, adamatzky_2009}.

We reported results of explorative, or scoping, experiments on colour recognition by slime mould.  The slime mould colour sensors are energy efficient: 
Physarum can survive on 1-2 oat flakes, 3~mm diameter each, for up to 5 days. The sensors are more likely to be used in application 
areas where a speed, or operating frequency, is not too high and where low cost is important.  These sensors could be integrated in hybrid bio-computing devices for image processing and computational geometry~\cite{shirakawa, adamatzky_ppl_2008, jones_2013} and as parallel optical inputs arrays in large scale Physarum computing devices~\cite{adamatzky_physarummachines, adamatzky_phychip}.  Physarum sensors can be also used 
in implementations of cheap disposable electronic circuits. Physarum sensor, and associated living slime mould circuitry, can be insulated with 
octamethylcyclotetrasiloxane  (Silastic 4-2735 Silicone Gum, Dow Corning S.A., B-7180 Seneffe, Belgium). Physarum remains alive and functioning 
while covered in insulator for days~\cite{adamatzky_wires}.

There plenty of work left to do to implement a fully functional and reliable colour sensor made from living   \emph{P. polycephalum} and to integrate Physarum colour sensors into hybrid wetware-hardware computing circuits made of the slime mould. Main issues to resolve are integration with silicon hardware, increase of reliability and maximising life-time.  Further integration of slime mould with silicon devices should not pose a big problem. Previously we demonstrated that Physarum can thrive on a very wide range of substrate from polymers to aluminium foil to electronic boards~\cite{adamatzky_physarummachines}. To increase reliability and repeatability of sensor outputs we can geometrically constrain Physarum: by preventing branching of a protoplasmic tube connecting electrodes we reduce a level of noise in oscillatory patterns.  With regards to maximising life time of Physarum sensors two options could be explored. First option could be to interface Physarum sensors with micro-fabricated vascular networks to deliver nutrients and remove products of metabolism~\cite{shin_2004, Lucarotti_2013}. Second option could be to partly load Physarum, especially a protoplasmic tube connecting electrodes, with functional and conductive nanoparticles~\cite{mayne_2013}.

\end{document}